\documentclass[]{spie}  

\usepackage[]{graphicx}

\title{A high-altitude, station-keeping astronomical platform} 

\author{Robert A. Fesen\supit{a} 
\skiplinehalf
\supit{a}Department of Physics \& Astronomy, Dartmouth College, 6127 Wilder Lab, Hanover, NH USA 03755
}

\authorinfo{E-mail: fesen@snr.dartmouth.edu} 

\begin{document} 
\maketitle 

\begin{abstract} 
Several commercial telecommunication ventures together with a
well funded US military program make it a likely possibility that an
autonomous, high-altitude, light-than-air (LTA) vehicle which
could maneuver and station-keep for weeks to many months will be a reality in a few
years. Here I outline how this technology could be used to develop a
high-altitude astronomical observing platform which could return 
high-resolution optical data rivaling those from space-based
platforms but at a fraction of the cost.  
\end{abstract}



\section{INTRODUCTION}
\label{sect:intro}  

There is currently considerable interest in the telecommunications industry and
the US military in developing an autonomous, high-altitude, light-than-air (LTA)
vehicle that could maneuver and station-keep for weeks to many months.  Several
telecommunication companies located both in the US and abroad along with
several branches of the US military are investing tens of million of dollars in
the development of these so-called hybrid airships: LTA vehicles consisting of
helium filled superpressure balloon(s) that are made maneuverable by means of
specially designed high-altitude propellers and powered by solar panels during
the day and solar charged batteries and/or power cells at night. If in the next
few years such vehicles are successfully built and flown, then these technologies
allow for the potential development of a high-altitude astronomical observing
platform which could return optical and infrared data rivaling
those from space-based platforms but at a fraction of the cost.

   \begin{figure}
   \begin{center}
   \begin{tabular}{c}
   \includegraphics[height=10cm]{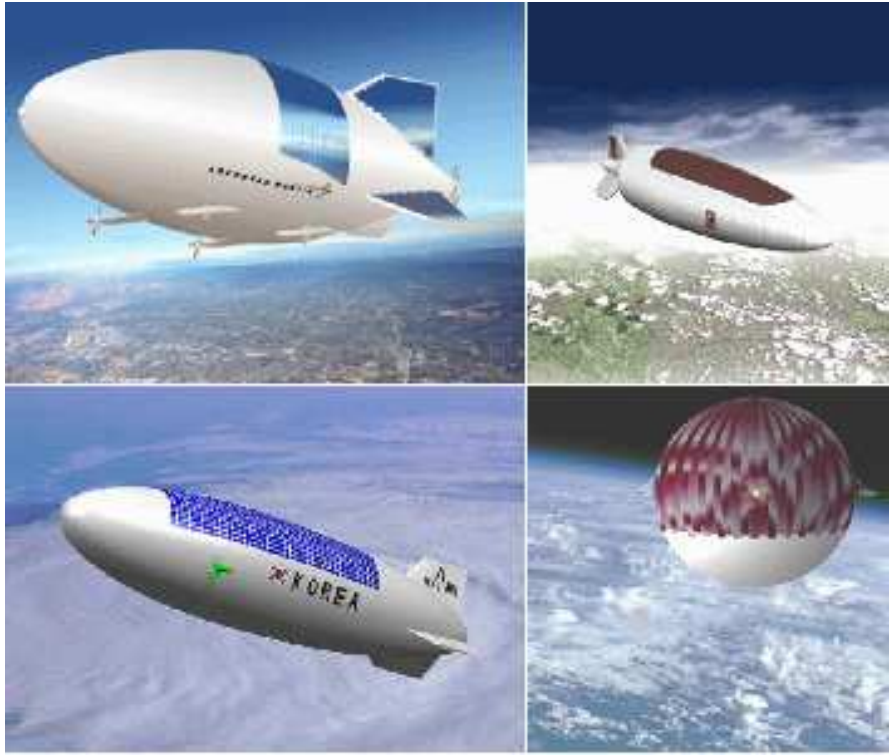}
   \end{tabular}
   \end{center}
   \caption[example]
   { \label{haa}
Artist concepts of the US Missile Defense Agency's high-altitude 
airship (upper left; Source: Lockheed Martin),
Lindstrand's high-altitude airship design for ESA (upper right; 
Source: Lindstrand Balloons Ltd),
Korea's HAP airship (lower left; Source: Korea Aerospace Research Institute), 
and a proposed spherical, high-altitude communications platform 
(lower right; Source: 21$^{\rm st}$ Century Airships).}
\end{figure}

\subsection{Commercial High-Altitude Telecommunication and Internet Platforms}

For more than a decade, wireless communications using a quasi-stationary,
high-altitude platform station (HAPS) for broadcasting services has been
considered by several telecommunication companies as a way to economically
expand commercial high-bandwidth data services to consumers.
Proposed HAPS vehicles have included high-altitude aircraft and LTA airships
operating at altitudes typically between 17 and 22 km  (55 -- 72 kft) where
stratospheric wind speeds are generally lowest. Companies and governments which
have pursed or are currently pursuing this technology using airships include
the US-based firms SkyStation, Platforms Wireless International, and Sanswire Networks,
the UK firm Advanced Technology Group (ATG), SkyNet and the Wireless Innovation
Systems Group of the Yokosuka Radio Communications Research Center in Japan,
and the Korean Ministry of Commerce, Industry, and Energy. The advantages and
technical problems of HAPS for wireless communications has been well 
laid out and discussed in several journal and magazine
articles\cite{Tozer01,Platt99,Djuknic97}.  While no full-scale autonomous vehicle has yet
been successfully flown achieving commercial payload and station-keeping duration
needs, several prototypes have been built and undergone successful initial test
flights. 

\subsection{`Near-Space' Military Reconnaissance and Tactical Platforms }

Across the military services, there is increased demand for real-time
communications and `over-the-horizon' surveillance capabilities. Recent advances
in LTA vehicles that operate at high-altitudes and hence offer large
surveillance areas and good air defense survivability factors have attracted
considerable interest from several military organizations.  For example, an
airship at an altitude of 70 kft would have a line-of-sight regional
coverage some 325 miles in diameter, meaning that just one such vehicle could survey
nearly all of Afghanistan.  Military interest also arises from the potentially
much longer on-station times of high-altitude airship platforms compared to
high-altitude UAVs (i.e., weeks to months vs.\ Global Hawk's $\sim$ 60 hours)
along with a low probability of communication intercept due to stable, direct
line-of-sight communications\cite{Jamison05}.  In this way, an airship would
function as surrogate satellite but offer shorter transmission distances in
theater battlespace and shorter ranges and hence higher resolution for sensor
surveillance of ground targets\cite{Jamison05}.

Probably the most aggressive and well funded military effort to develop a
high-altitude airship (HAA) has been pursued by the US Missile Defense Agency
(MDA) through its Advance Concepts Technology Development (ACTD) program.  This
program's goal is the development of an autonomous airship that would be
stationed at 65 kft, have an endurance of at least one month (with one year
desired), carry a payload of 4000 lbs and be capable of supplying that payload
with 10 kW of power. The vehicle envisioned would have a top airspeed of over 30
knots to station-keep on a designated ground target and be some 450 ft long and 150 ft
in diameter with over 5 million cubic feet in volume. Lockheed Martin's Naval
Electronic and Surveillance Systems (Akron OH) was selected for the initial
design study and awarded US\$40M in September 2003. It was recently awarded
US\$149M in December 2005 to begin building a small prototype (top speed 25 kts, 500 lb
payload with 3 kW of power) to fly before 2010. The general vehicle concept for
the Lockheed Martin design is that of a conventional dirigible shape with
photovoltaic cells across the upper surfaces (see Fig.\ 1). Station-keeping
target parameters are $<$ 2 km for 50\% of the time and $<$150 km 95\%
of the time.

The US Air Force is also interested in an unmanned vehicle to loiter at 100 kft
for several days carrying a 50 kg surveillance payload and awarded a contract
to JP Aerospace through the Space Battlelab and Space Warfare Center in
Colorado.  The European Space Agency (ESA) has also considered development of a
solar-powered airship and awarded a design study to Lindstrand Balloons (see
Fig.\ 1). Their airship design is similar to the MDA airship but with a single
8-meter diameter propeller and 15 kW of power for a communications payload.

\section{Science Platforms at the Edge of Space}

The idea of developing a high-altitude science platform for astronomical
studies is not new, dating back more than a century.
Astronomers have long realized that stellar scintillation and other atmospheric
effects could be greatly eliminated by placing a telescope at stratospheric
altitudes.  Attempts at high-altitude astronomical observations may have 
begun with James Glaisher's balloon ascents in the 19th century, followed
much latter by A.  Dollfus's ascent to 42 kft in 1959 to observe the planet
Venus and the Air Force's Cambridge Research Lab's {\sl Project Stargazer} in
1961\cite{Ryan95}.  

   \begin{figure}
   \begin{center}
   \begin{tabular}{c}
   \includegraphics[width=14cm]{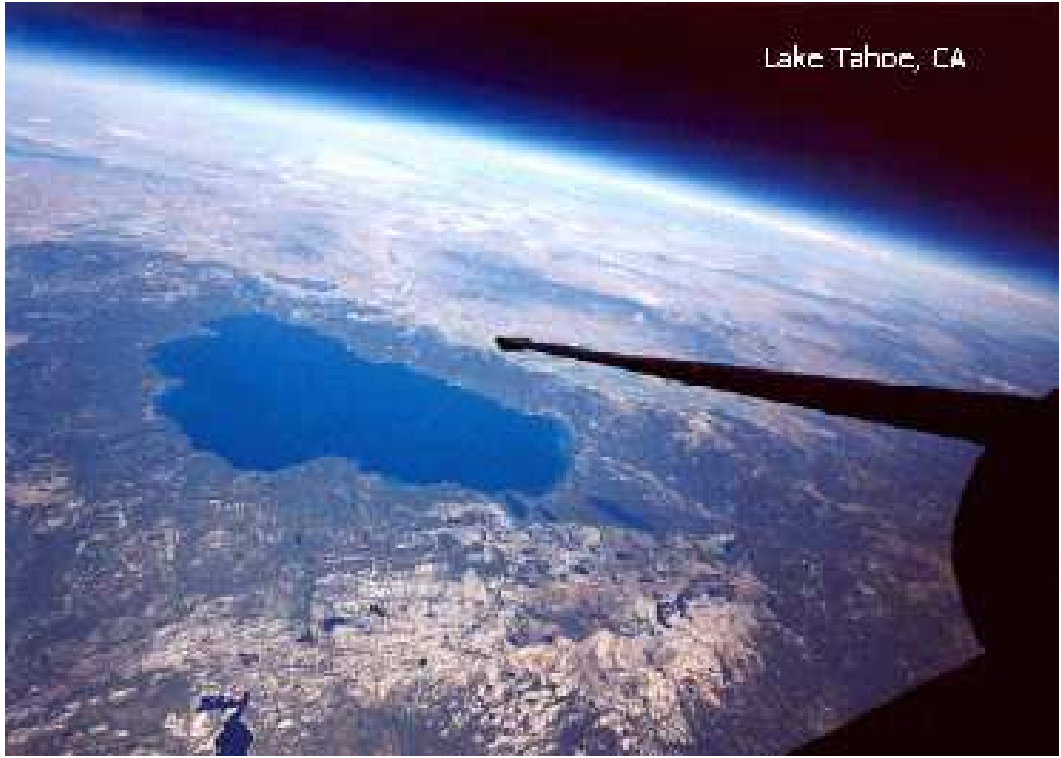}
   \end{tabular}
   \end{center}
   \caption[example]
   { \label{U2_image}
How high does one need to go to avoid clouds and stormy weather and begin obtaining
space-like astronomical observing conditions? The answer is between 65 kft and 85 kft
(see text). Here's a photo taken from a U2 aircraft flying over Lake Tahoe CA at an altitude
of 70 kft.  Note the curvature of the Earth along the horizon and
the very dark sky even during daytime. The U2's wing tip and the outboard section of the mid-wing
sensor pod can be seen in the right-hand side of the picture. (Source: US Air Force)
}
\end{figure}

Little useful science was produced by these early efforts.  However, this
changed when Martin Schwarzschild's {\sl Stratoscope} balloon carried a 12-inch
refractor to 80 kft and returned the first high-resolution images of the Sun despite
image stabilization problems, and later when the {\sl Stratoscope II} balloon
carried a 36-inch infrared telescope to observe Mars\cite{Ryan95}.  More recent
uses of high-altitude balloons for astronomical observations have been concentrated
in the X-ray, infrared, and millimeter regions of the electromagnetic spectrum. Some of
the more notable recent high-altitude balloon missions include 
BOOMERanG and GLAST.

NASA's Long Duration Balloon (LDB) and Ultra-Long Duration Balloon (ULDB) vehicles
are an attempt to provide high-altitude science platforms for missions that would
benefit from longer stays at high float altitudes. However, due to `no-fly' restrictions
of many nations in both northern and southern hemispheres, 
LDB and ULDB flights have largely been limited over Antarctica and Australia.
Consequently, a couple of high-altitude, quasi-station-keeping astronomical platforms have
been proposed within the last 10--15 years including the tethered Polar Stratospheric Telescope
(POST). 

Despite considerable interest by the atmospheric and Earth
science communities in developing a HAP similar to that sought after by
the telecommunications industry and the US military, there has been little interest
by astronomers to consider a HAP for optical and near-infrared astronomical
observations. This may be due in part to the availability and great success of the
{\sl Hubble Space Telescope} (HST) and the almost exclusive use of previous
balloon missions for X-ray, infrared, millimeter, and cosmic ray observations.  One notable
exception to this is the planned high-altitude 1 m solar telescope project {\sl
SUNRISE}. 

While no high-altitude, station-keeping science platforms currently exist, over
the past several decades there have been several attempts by various
organizations. Unfortunately, many of these were poorly funded or
executed\cite{Smith03}.  A common mistake was attempting to progress too
quickly from a small prototype to lifting large payloads very high up using
huge experimental balloon vehicles. This often lead to failure and program
termination\cite{Smith03}.  However, two vehicles that did come close to the
goal of a maneuverable high-altitude platform was {\sl HighPlatform II}
developed by Raven Industries in the late 1960's, and the Army funded {\sl
Sounder} project by SWRI in the late 1990's. This latter program was used as a
baseline design vehicle as a potential science platform in a NASA project
funded in 2000 through the Cross Enterprise Technology Development Program.  


\section{A High-Altitude Astronomical Observatory} 

Despite an ever increasing number of scientific satellite missions, 
there are strong science and financial drivers for development of a relatively
inexpensive, aerial vehicle for a wide range of Space and Earth Science 
applications. For example, at an altitude of 85 kft an astronomical telescope would
experience virtually perfect skies overhead every night with image quality approaching
the diffraction limit of the main aperture. An optical telescope with a
lightweight mirror just 20-inch in diameter (0.5 m class) with sufficient
pointing stability and large CCD arrays could provide wide-field images with 
FWHM = 0.25 arcsec, making it superior to the imaging system on any ground-based telescope.
And it could do it night after night for as long as the platform
remained at this altitude.  Moreover, such a stratospheric telescope could
also provide reliable science support for a host of space-based missions at an
estimated cost of a few percent of a conventional low Earth orbit (LEO) satellite.

Below, I discuss the potential use of a high-altitude airship for use as a
scientific observing platform, specifically an astronomical imaging observatory. A
stable aerial platform positioned at low- to mid-stratospheric altitudes would
provide a space-like observation outpost far more accessible and less expensive
than satellites.

\subsection{Advantages of a High-Altitude Airship for Astronomical Observing}

One of the chief advantages of a near space, stratospheric location is the
greatly reduced atmospheric effects on astronomical image quality.  At an
altitude of 65 kft (20 km) one is above all but 5.5\% of the atmosphere,
whereas at 85 kft (26 km) less than 2.3\%  of the atmosphere lies overhead.
Just as important is the lack of appreciable water vapor and dust and other particulates  in
the remaining atmosphere above these altitudes.  As can be seen in Figure 2,
there is no weather up at these altitudes, with 60 kft being about as high
as the tallest thunderstorms ever get.  Pilots operating high altitude aircraft
through the 70--85 kft regime, such as the U2 and SR-71, report remarkably smooth
flying conditions throughout most parts of the world\cite{Shul91}.

In this thin, dry and clear atmosphere, optical systems can approach their
diffraction limit much like that on space vehicles. Although little in the way
of hard image quality numbers vs.\ stratospheric altitudes are known to the author, the
evidence available supports the likelihood of near-space seeing conditions.  Figure 3
shows plots of the predicted coherence length, r$_{\rm o}$, and the isoplanatic
angle with altitude (L.\ Petro, priv.\ comm.). The coherence length is
effectively the largest diameter of a lens or mirror over which the phase of
the incoming light waves is coherent, thereby permitting diffraction limit
imaging.

   \begin{figure}
   \begin{center}
   \begin{tabular}{c}
   \includegraphics[width=17cm]{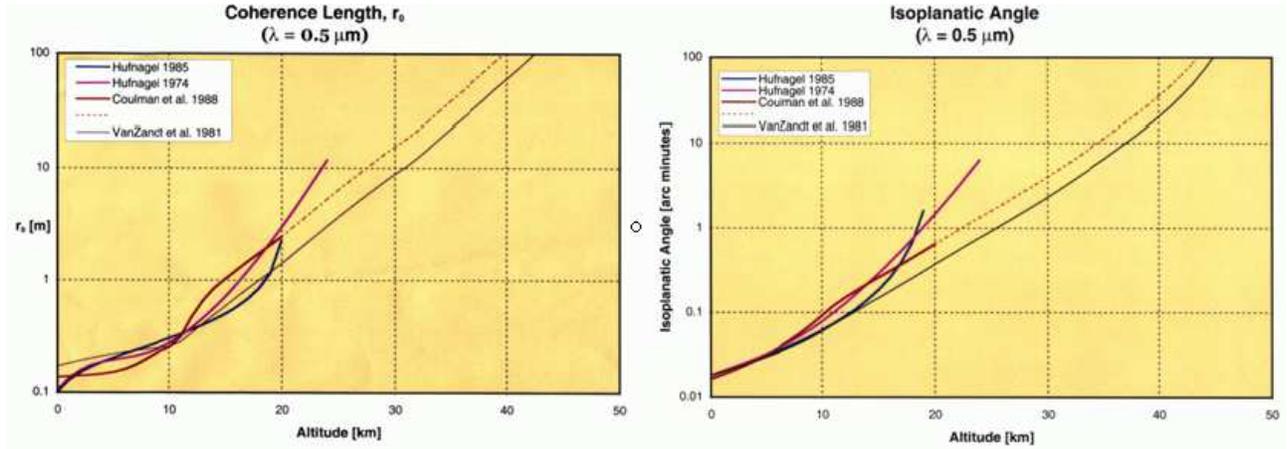}
   \end{tabular}
   \end{center}
   \caption[example]
   { \label{seeing_plots}
Theoretical model plots for the coherence length (left) and the isoplanatic angle (right).
(Source: L. Petro, STScI) }
\end{figure}

On the ground, the coherence length is typically about 6 inches (0.15
m) across. As shown in the left panel of Figure 3, the coherence length is predicted
by several models to increase steadily from ground level, reaching 1 meter at altitudes between
13--17 km (40--52 kft). At an altitude of 25 km (82 kft), the
value of r$_{\rm o}$ is predicted to be in excess of 2 m and could be as large
as 10 m. This means that at stratospheric altitudes like those planned for
telecommunication and military HAAs, an astronomical telescope with a mirror 2
meters or greater in diameter could, in principle, return diffraction limited
images approaching that of HST.  

The isoplanatic angle, i.e., the angle over which a diffraction limited image
can be instantaneously obtained, is shown in the right panel of Figure 3 as a
function of altitude for visible light ($\lambda$ = 500 nm).  The plot shows that the
isoplanatic angle is expected to be in excess of 1 arcmin at 25 km and could be
as large as 10 arcmin.  Due to air flow and turbulence, the isoplanatic patch
will move across the mirror causing the image to shift across a detector. The result is 
that a 2-meter telescope with diffraction limited optics will need 
superb image tracking and stabilization to yield near space-like
images (FWHM = 0.0625 arcsec at 500 nm).
 
Other advantages of an astronomical HAA include no ground-site to be purchased
or developed, and little light pollution if flown far away from populated
regions. Due to absence of an appreciable atmospheric column of air, water
vapor and dust overhead, excellent horizon-to-horizon observing is possible
if not obstructed by the airship itself. In addition, light scattering from
moonlight would be expected to be minimal and not a factor in scheduling faint
target observations, with most observing time effectively `darktime'.  This
would greatly enhance the platform's ability for rapid response for
observations of faint transient targets such as supernovae and gamma-ray
bursters.  Also, unlike LEO satellites such as HST, data transfer to/from a HAA
could involve simple line-of-sight communications running 24/7.  The ground
control and operation station could be sited in a location some distance away
from the airship's geographical overflight location. 

\subsection{Engineering and Environmental Obstacles} 

There have been many studies dealing with various engineering issues regarding the
development  and successful flight testing of high-altitude hybrid
platforms\cite{Tozer01,Jamison05} and I will not discuss these in detail here. 
However, major technology and
performance issues for airships include engine power, envelope fabric strength
and UV durability, fuel-cell capacity, and launch and recovery procedures.  Of
these, the durability of airship fabric under intense UV radiation and high
ozone levels at these altitudes are particularly important problems for long
term observatory flight times. 

On the other hand, some issues facing telecommunication and military airships are not as important
in the development of a science observatory. These include strict
station-keeping limits over particular ground targets and payload recovery
schemes that do not involve populated areas or loss of sensitive equipment of a defense
or national security nature. The high power requirements (i.e., many kW) for military and
telecommunication payloads is also not expected to be a problem for an
astronomical telescope and CCD detector system. 

Perhaps the two greatest challenges for a high-altitude astronomical facility
are keeping the science payload weight down so as to require only
a relatively small and maneuverable airship and a lightweight but robust telescope pointing
system that can acquire and track targets with great precision ($< 0.05$
arcsec). Lightweight mirror technology has made great strides through programs
such as the Next Generation Space Telescope (NGST) and several lightweight
mirror materials (e.g., beryllium, silicon carbide, and graphite fiber
composites) have been showed to yield good results using active wavefront
controls\cite{Takeya04,Angel04,Twarog04,DeV05}. Image stabilization and target
tracking, however, may be a bigger problem given the severe payload weight
restrictions of a maneuverable LTA platform. 

\section{Recipe for a High-Altitude Astronomical Observatory}

Below, I offer some guidelines on how a HAA might be designed,
where geographically and at what altitude might be best to
operate at, and discuss some issues regarding instrument options.
Many of these recommendations are from the findings of a NASA sponsored
study of a scientific HAA in 2000 at the Goddard Space Flight Center.

\subsection{Basic Airship Design}

A key ingredient in the design of an astronomical platform is the ability of
the telescope to survey most if not all of the sky (from horizon-to-horizon),
ideally without obstruction from the vehicle itself.  The telescope should also be placed on
the vehicle in a such a way that it can track objects regardless of the
airship's orientation or direction of motion. Moreover, the vehicle design
should be such that it provide a good degree of inherent stability to help facilitate
high-resolution imaging over timescales of minutes to hours.

The resulting technologies and design solutions derived from the various
telecommunication and military efforts now underway will, of course, determine
much of the overall structure of a successful HAA.
However, such commercial and military airships are designed
to carry fairly heavy payloads primarily on the lower surfaces of the vehicles,
i.e., downward looking, and thus not likely to be directly suitable for
astronomical purposes.

   \begin{figure}
   \begin{center}
   \begin{tabular}{c}
   \includegraphics[width=17cm]{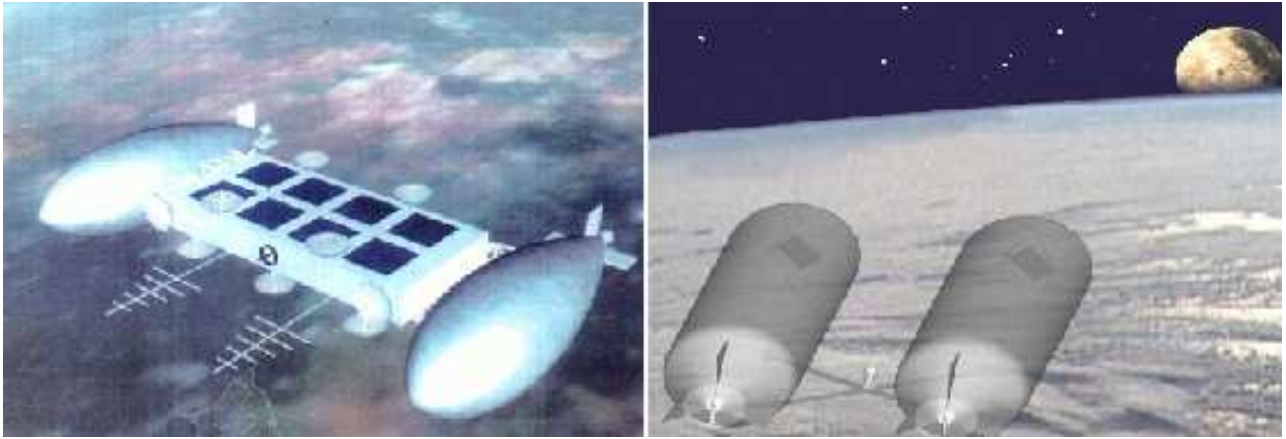}
   \end{tabular}
   \end{center}
   \caption[example]
   { \label{catamaran_designs}
Examples of proposed catamaran designs. Left panel: Proposed superpressure catamaran airship
for communications. (Source: Dr.\ A.\ Wong, Platform Communications Corp.) Right panel:
Artist concept for a catamaran airship design for an astronomical observatory using twin
Sounder-like hulls, with a small telescope mounted in between. (Source: W.\ Perry, SWRI) }
\end{figure}
                                                                                                                                                                            
   \begin{figure}
   \begin{center}
   \begin{tabular}{c}
   \includegraphics[width=17cm]{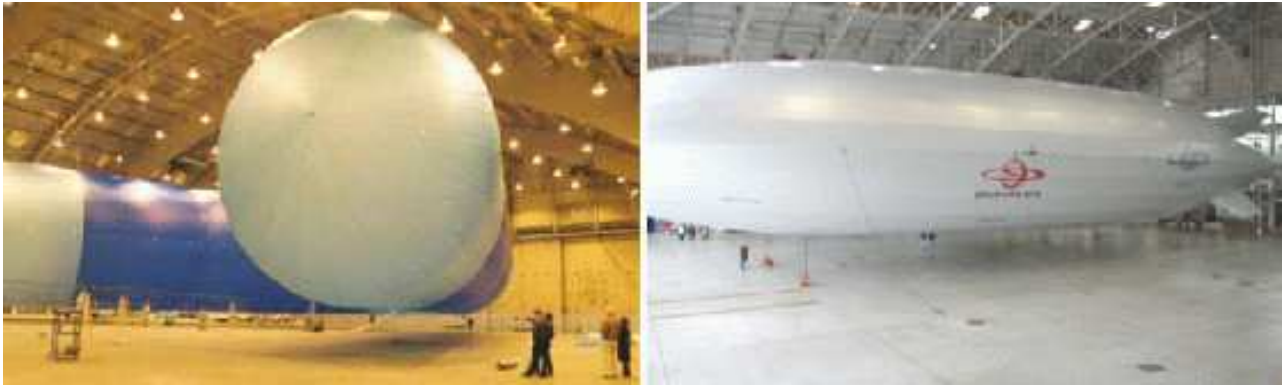}
   \end{tabular}
   \end{center}
   \caption[example]
   { \label{airships_cats}
Constructed airship prototypes of a size that might be used to carry a small telescope  
to high altitudes. Left: V-shaped {\em Ascender} 
(Source: JP Aerospace); right: A `stratellite' {\em Sanswire One} (Source: Sanswire Networks).
}
\end{figure}
                                                                                                                                                                            
One possible HAA design that has received considerable attention uses two
superpressure balloons arranged in a catamaran fashion. Figure 4 shows examples
of previously proposed catamaran HAAs, one for telecommunication purposes and
the other for astronomy. In both cases, the payload is mounted in-between two
cylindrical balloons. Of course, unlike that shown in the twin {\em Sounder}
catamaran design (Fig. 4, right panel), mounting the telescope high up would
gain the ability to see over the rest of the vehicle and hence have a clear
and unobstructed view of the sky. Heavier components of the HAA systems,
e.g., batteries, power cells, communications hardware, and possible even
some of the propulsion system might be hung from below the HAA to
provide stability off-setting the placement of the telescope high up on the
platform.  In any case, the development of an astronomical platform will need to
follow the basic design components of the successful commercial and military vehicles 
(see Fig.\ 5) and then
adapt these to the goal of carrying a lightweight telescope that can
rapidly slew to and track astronomical targets without re-positioning the
airship.

\subsection{Operational Altitudes}

Nearly all station-keeping HAAs have been designed to operate in the 
60--70 kft altitude range, where average wind speeds tend to be the lowest (5--10 
knots; 10--20 m/s: see Fig. 6). The 60--65 kft altitude range is sometimes
referred to as the ``sweet spot'' for HAAs, because at this altitude one is above nearly all storm
systems yet with sufficient air density to give good engine efficiency
making it easier to move large airships. Going from 60 kft to 100 kft
doubles the horizon-to-horizon coverage, which is great for ground-sensing and
telecommunication purposes, but wind gusts at this altitude can seasonally
reach 70--150 kts (35--70 m/s). Since most HAAs are designed to have daytime top speeds
in the 15--25 m/s range, this would preclude reliable station-keeping for certain times
of the year. 

   \begin{figure}[b]
   \begin{center}
   \begin{tabular}{c}
   \includegraphics[height=5cm]{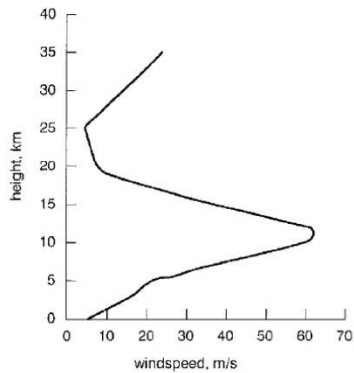}
   \end{tabular}
   \end{center}
   \caption[example]
   { \label{winds_aloft}
Typical wind speed vertical profile through the atmosphere. (Source: National Weather Service)}
\end{figure}

Typically, the highest stratospheric winds are seen at mid- to polar-latitudes,
with the lowest winds seen near the equator.
This is shown in the two plots shown in Figure 7.  
Whereas the military and communication firms are designing airships
that need to station-keep at mid-latitudes fighting at times very high winds,
a high-altitude astronomical observing
need not have that latitude restriction. Instead, a location near
the equator offers generally lower stratospheric wind speeds over
much of the year. As the right panel illustrates, although wind gusts
even at the equator can be as high as 40 m/s, winds are less than about 20 m/s about 90\%
of the time.

However, these equatorial wind speed plots do not tell the whole story.
There is a well studied atmospheric phenomena in the tropics known as the
quasi-biennial oscillation (QBO) that is seen as abrupt wind direction changes with a
period around 2.3 years.  This is shown in Figure 8 where one sees rapid east
and west wind changes (i.e., positive and negative winds speeds). This figure
shows that the winds are much faster at 90 kft vs.\ 67 kft where  the
average peak wind speeds are 10 m/s or less. Hence, a HAA located near the
equator and at a float altitude of around 70 kft would offer relatively low
stratospheric wind speeds throughout the year, thereby reducing the
propulsion power required to station-keep.

   \begin{figure}
   \begin{center}
   \begin{tabular}{c}
   \includegraphics[height=5cm]{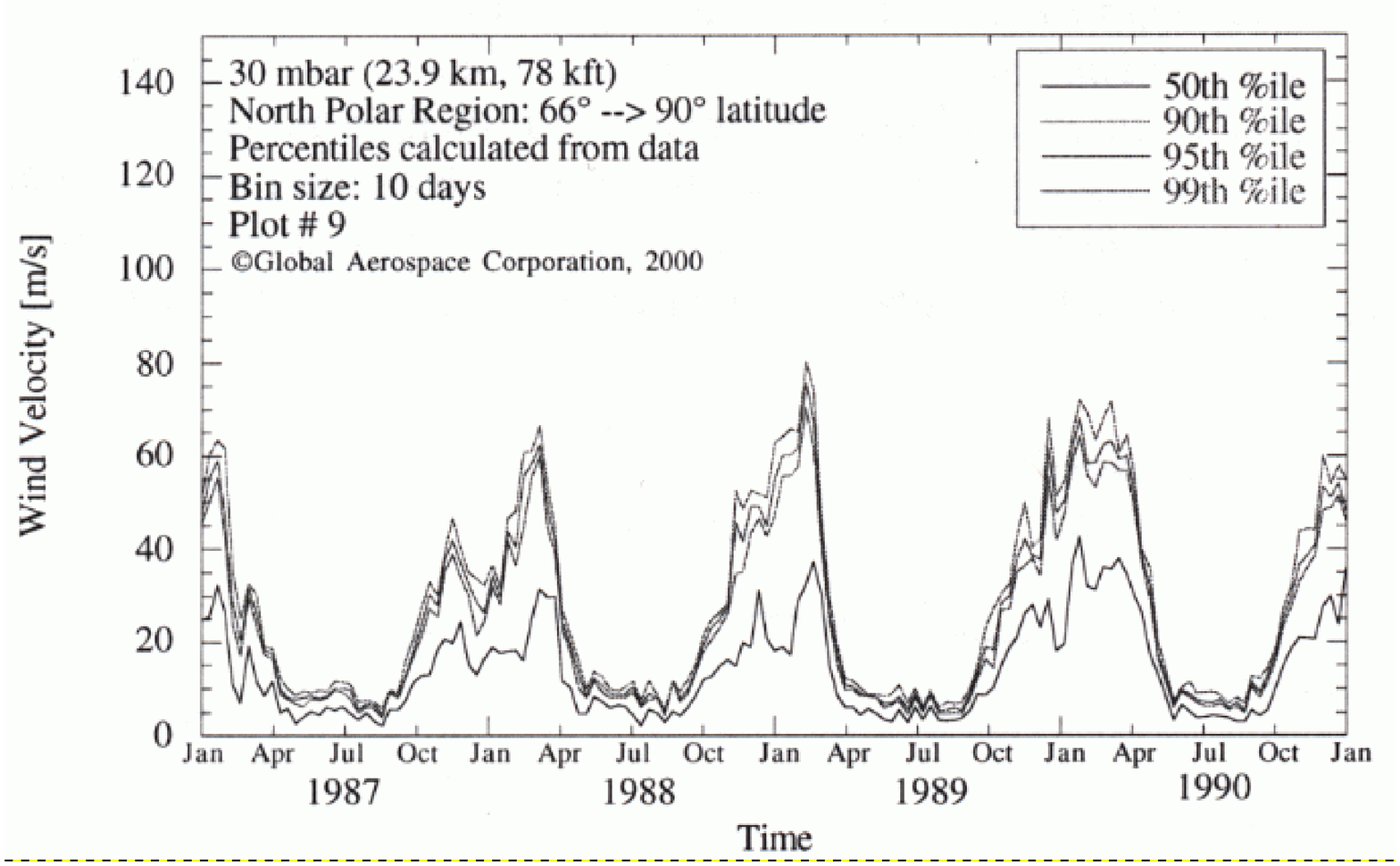}
   \includegraphics[height=5cm]{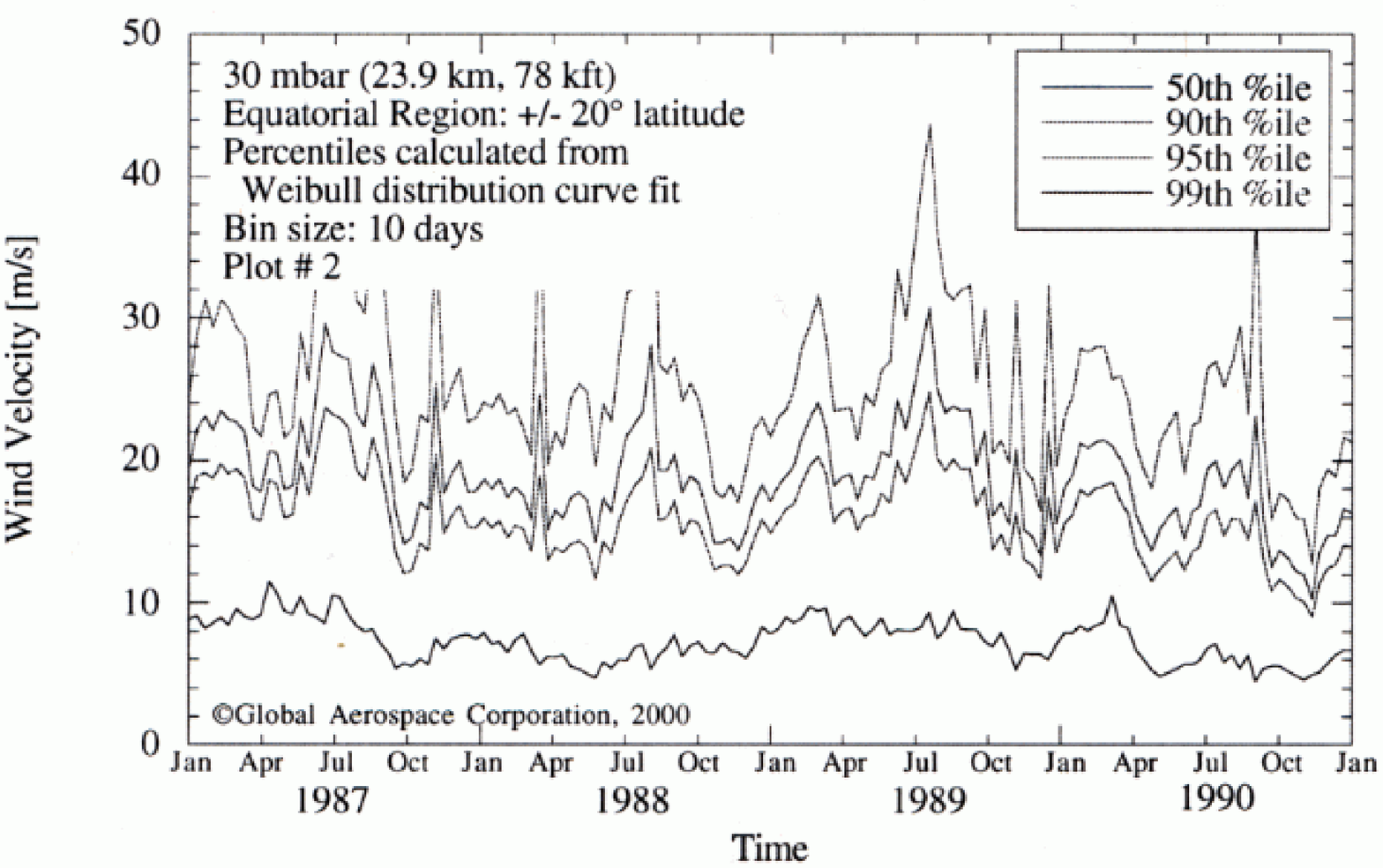}
   \end{tabular}
   \end{center}
   \caption[example]
   { \label{fig:Global_aero}
Seasonal variations of stratospheric wind speeds at 78 kft (30 mbar)
for mid to high latitudes (left panel) and for the equatorial
region namely, latitude $\pm$ 20$^{\rm o}$. (Source: Dr.\ Matthew Heun,
Global Aerospace Corp.)  \\
~
}
\end{figure}
   \begin{figure}
   \begin{center}
   \begin{tabular}{c}
   \includegraphics[height=5cm]{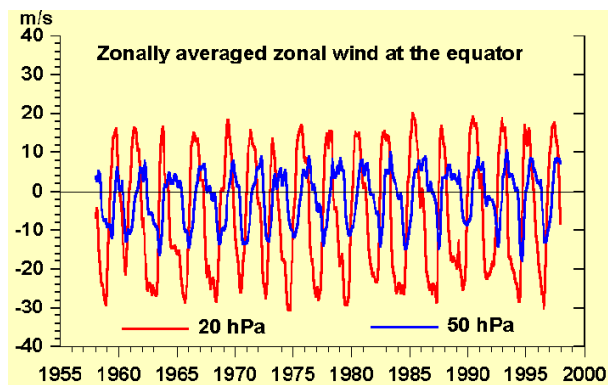}
   \end{tabular}
   \end{center}
   \caption[example]
   { \label{QBO}
Monthly means of zonally averaged zonal winds at 
the equator for 20 hPa (90 kft) and 50 hPa (67 kft) pressures (altitudes) during 1958-1997.
(Source: A. Kapala and H. M\"achel (NCAR))}
\end{figure}

\subsection{Geographic Location}

Interestingly, the equator also offers an excellent location on scientific grounds
for a high-altitude astronomical observing station. Located at or near the
equator, a telescope on a HAA would have access to astronomical targets
in both the southern and northern hemispheres. At an altitude of 70 kft,
such a platform would be above 95\% of the atmosphere and above 99+\% of the
water vapor. Hence, atmospheric extinction would be minimal even near the
horizon, making true horizon-to-horizon observations possible. This would not only
increase the number targets visible during a given night, but also the duration
over which they could be observed;, i.e., virtually from the time they rise above the horizon to
when they set. The platform would also experience nearly ideal photometric conditions night after night,
meaning a sequence of long-term observations could be linked together much like that
done using space-based data sets.

There are many islands in the Pacific (including several with US military
bases) which could be used for operating such a vehicle.  Platform operation
over open ocean would also lessen launch and recovery safety
issues, catastrophic vehicle malfunction fallout problems, international no-fly zones, and
interference with commercial and military air traffic. 
An ocean location would also afford little or no artificial light pollution.

It turns out that an ocean location near the equator would also eliminate two problems
with HAAs that are significant concerns for operations over land
and at mid-latitudes. One is the near absence of hurricanes and cyclones
occurring within 10 degrees of the equator. While hurricane cloud tops do not
typically reach altitudes above 60 kft, storm downdrafts can effect
balloons floating as high as 70 kft, pulling them down to much lower altitudes as was demonstrated
early on in the US military {\em Manhigh II} balloon flights\cite{Ryan95}.  In addition,
broad (5--30 km) and powerful upward lightning discharges from the tops of violent
thunderstorms into the stratosphere (altitudes: 50--90 km) known as `sprites'
are mainly limited to particular geographic regions such as Florida and
the US Midwest and rarely occur over open ocean.

\subsection{Instrumentation}

The principle limitation for any astronomical instrumentation on a
high-altitude LTA observing platform is weight. The heavier the payload, the
bigger the airship is needed to lift that payload and the more power required
to push the airship to station-keep against the ambient winds. The
mirror, the optical telescope assembly, filters, filter wheels, and the
detector package must all be made as light as possible while still delivering high
quality science data. In addition, an  ideal instrument should not require any
expendables such as LN2 to cool its detector.  

The simplest and lightest weight instrument and detector system that could
generate first-rate science results is that of a wide-field CCD imager. A single $4096
\times 4096$ pixel CCD mounted at prime or Cassegrain focus on a airborne 0.5 m
telescope could yield images with a field of view of $6.5' \times 6.5'$ with an
image scale of 0.1 arcsec pixel$^{-1}$ and a diffraction limited resolution of 0.25
arcsec.  Because the ambient air temperature at 70 kft is around $-55$ C, only
a modest amount of thermoelectric cooling might be needed to lower the CCD
temperature to the typical $-80$ to $-100$ C astronomical CCD operating
range.

In a 2000 NASA HAA study, preliminary designs
based on a payload weight of 300 kg proved quite challenging leading to
subsequent studies with smaller 50 and 100 kg payload weight limits. The Lockheed Martin HAA
may have undergone a similar progression of moving from heavy to lighter
payload goals in formulating the MDA prototype design vehicle. As Smith and
Rainwater warn: large payloads, high altitudes, and long duration
requirements are key components to program failure\cite{Smith03}. Therefore the simplest,
lightest instruments that can still yield valuable data should be considered as
initial pathfinder instruments before larger and more complex instrumentation
(such as a spectrograph) are attempted. The high precision photometry of bright
stars using the 2-inch star tracker on the {\sl WIRE} satellite is an example
of a very small telescope generating valuable science data.

It is perhaps in the optical regime that a HAA observatory might be
most effective. Adaptive optics (AO) techniques can
produce near diffraction-limited image from ground-based telescopes
in the infrared. However, using either natural or artificial stars 
generated by lasers, this technique affords a very limited field of view (FOV),
($<$ 30 arcsec) with high-resolution. Moreover, AO is unlikely to be effective
at optical wavelengths (i.e., $<$ 1 micron) within the next several years.
Space observatories like the {\sl Hubble Space Telescope} can provide
high-resolution optical images but over a relatively small FOV ($\sim 200'' \times 200''$).
HST also has severe restrictions on the number of target of opportunity requests
that require fast acquisition times. In both these regards, a high-altitude
astronomical telescope equipped with even an amateur-sized telescope mirror could
be an effective tool for wide-field imaging where sub 0.1 arcsec resolution
is not needed but steady Strehl ratios are, or cases where fast imaging response times are desired.

Finally, although there are clear opportunities for optical imaging, there are
also significant advantages that a high-altitude observatory offers in the
infrared if a suitable detector is used which can operate effectively without the need
of LN2 or liquid helium.  Figure 9 compares the atmospheric transmission across the
optical and infrared wavelength regimes for a high-altitude, ground-based
observatory (Mauna Kea, HI; altitude; 14 kft: 4.2 km), NASA's airborne IR
observatory {\sl SOFIA} (expected cruise altitude = 40 kft: 14 km), an
astronomical HAA as discussed above but at its uppermost likely altitude range (91 kft:
28 km), and an ultra-high-altitude science balloon (135 kft: 41 km).  

This figure shows that although a large improvement occurs in the atmospheric
transmission going from a mountain top location to the maximum cruise
altitude of a commercial jet aircraft ({\sl SOFIA}: Boeing 747-SP), there is a large
additional improvement moving up from 40 kft to 90 kft.  In addition, it should
be noted that an HAA telescope 
will not be subject to high winds, especially if the airship is operated in a `sprint and
drift' day and night station-keeping mode. Just as important, cirrus
clouds which might affect photometric conditions at 40 kft and which cannot be easily
predicted or calibrated, are of no concern to an airship floating at altitudes
above 65 kft.  However, as shown in Figure 10,
some water vapor ($\sim10-30$ ppmv) is still present at stratospheric altitudes
at both the poles and equator. Thus, a HAA operated near the equator at a pressure level of 50 mbar (67
kft; 21 km) to avoid the highest QBO winds would not be above most of this
water vapor layer.

   \begin{figure}
   \begin{center}
   \begin{tabular}{c}
   \includegraphics[height=11cm]{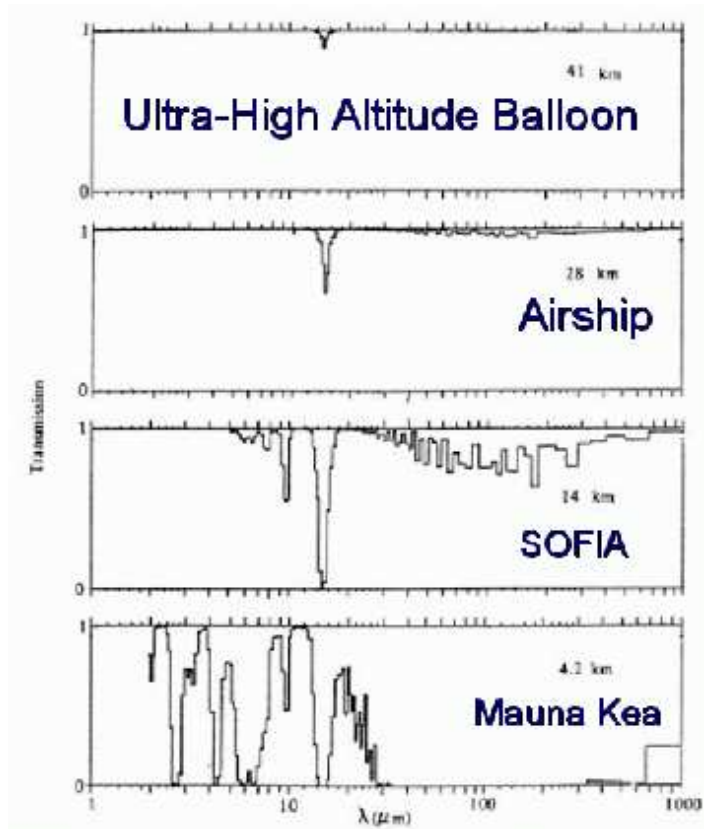}
   \end{tabular}
   \end{center}
   \caption[example]
   { \label{atm_trans}
Atmospheric transmission from the optical to the far infrared as a function of altitude.
Adapted from: {\em Handbook of Space Astronomy and Astrophysics}\cite{Zombeck} }
\end{figure}

   \begin{figure}
   \begin{center}
   \begin{tabular}{c}
   \includegraphics[height=9cm]{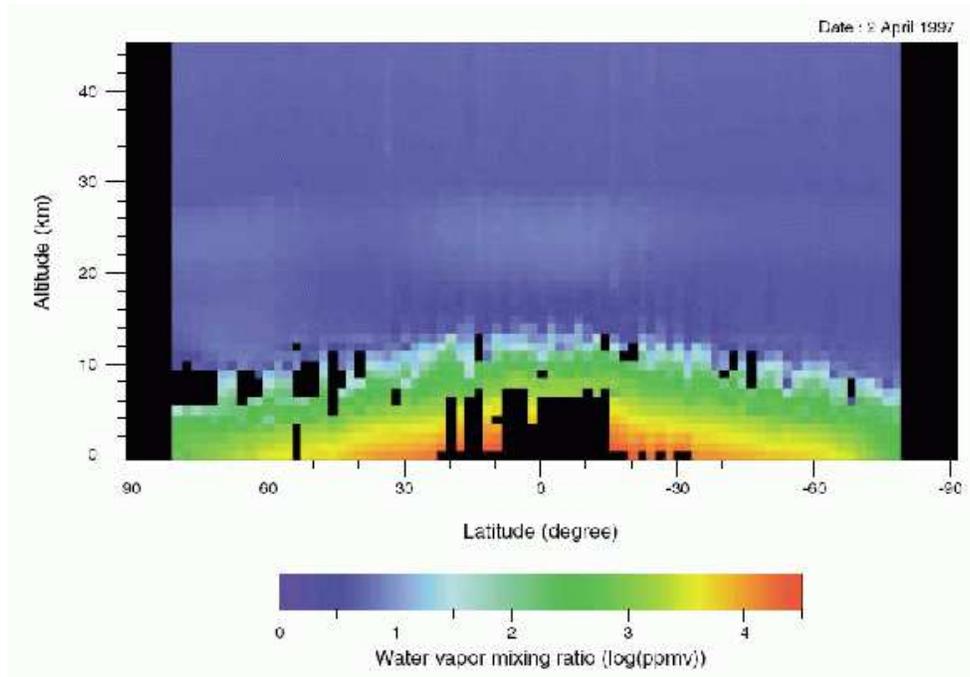}
   \end{tabular}
   \end{center}
   \caption[example]
   { \label{atm_example}
Map of atmospheric water vapor percentages (ppmv) as a function of both latitude and atmospheric altitude.
Image shows a band of water vapor over low latitudes
extending from 20 to 28 km in height. (Source: NASA)
}
\end{figure}
                                                                                                                                                                            
\acknowledgments     
 
The author would like to thank Bill Perry (SWRI) for many hours of stimulating 
discussions about high-altitude airships and his {\sl Sounder} design.


\end{document}